\documentclass[aps,pra,oneside,twocolumn]{revtex4-1}
\usepackage{amsmath}
\usepackage{amssymb}
\usepackage{graphicx}
\usepackage{multirow}
\usepackage{booktabs}

\begin{document}
\draft \title{Modeling off-resonant nonlinear-optical cascading in mesoscopic thin films and guest-host molecular systems} \author{Nathan J. Dawson, James H. Andrews, and Michael Crescimanno}
\address{Department of Physics and Astronomy, Youngstown State University, Youngstown, OH 44555}
\email{Corresponding author: dawsphys@hotmail.com}

\begin{abstract} A model for off-resonant microscopic cascading of (hyper)polarizabilities is developed using a self-consistent field approach to study mesoscopic systems of nonlinear polarizable atoms and molecules. We find enhancements in the higher-order susceptibilities resulting from geometrical and boundary orientation effects. We include an example of the dependence on excitation beam cross sectional structure and
a simplified derivation of the microscopic cascading of the nonlinear optical response in guest-host systems.
\end{abstract}

\maketitle

\section{INTRODUCTION}
\label{sec:intro}

Although of increasingly greater importance in compact optical devices and
metrology, structural mesoscopic nonlinear-optical effects have not been
extensively studied.
In bulk systems, it is well known that cascaded nonlinear optical interactions and local field effects at the molecular level can enhance higher-order nonlinear optical susceptibilities.\cite{andre91.01,stege93.01,stege96.01,baev10.01} Dolgaleva, \textit{et al.}, showed that local-field corrections predict trends in the nonlinear susceptibilities as functions of concentration in a bulk material.\cite{dolga07.01,dolga09.01,dolga12.01} Earlier studies focused on a tensor formalism to describe correlated cascading effects in bulk materials,\cite{andre91.01} while others focused on cascading between coupled molecules only.\cite{dawson11.02,baev10.01}  Mesoscale nonlinear optical effects, however, have not been well investigated, and give new insights into enhancing the nonlinear susceptibility that are not present in a bulk approximation.\cite{dawson12.01,*dawson12.01corr}

Here, we use the self-consistent field approach to cascading (Bloembergen's method \cite{bloem96.01}) to approximate the sum of the
dipolar response fields and thereby the
cascading contribution in mesoscopic systems.
We compute the effective (hyper)polarizabilities and susceptibilities
with respect to the applied field by an iterative update method to
approximate a finite ensemble of polarizable molecules. Of experimental
relevance, we apply this technique in realistic model systems to
quantitatively illuminate the role that boundaries and geometrical
orientation play in nonlinear susceptibility enhancement.

After describing our method in Section \ref{sec:firstorder}, we apply it in Section \ref{sec:appsing} to bounded and strained tetragonal systems. The dipolar field at each molecule from all other molecules is shown for different film thicknesses, where the dipoles are induced by a linear polarized Gaussian beam. Then, as an application to a real system, we find that the relationship between the cascading contribution of hexagonal close-packed and honeycomb structured monolayers of the molecule C$_{60}$ can be understood by the fill factor and concentration.  Finally, in Section \ref{sec:guesthost} we approximate the effective second hyperpolarizability of a mesoscale guest-host system in which a nonlinear dopant has been randomly distributed in a discretized linear matrix, providing an example of matrix-enhanced dye polarizability.

\section{THEORY}
\label{sec:firstorder}

\subsection{Self-consistent approach}
\label{sub:selfconsapp}

When point molecule $j$ is polarized by an electric field, it becomes a dipole, causing molecule $i \neq j$ to experience a corresponding dipole field
\begin{equation}
\boldsymbol{E}_{i,j} = \frac{3\left(\hat{r}_{i}-\hat{r}_{j}\right) \left[\boldsymbol{p}_{j}\cdot \left(\hat{r}_{i}-\hat{r}_{j}\right)\right] - \boldsymbol{p}_{j}}{\left|\boldsymbol{r}_{i}-\boldsymbol{r}_{j}\right|^3},
\label{eq:labframeE}
\end{equation}
where $\boldsymbol{p}_{j}$ is the dipole moment of molecule $j$,  $\left|\boldsymbol{r}_{i}-\boldsymbol{r}_{j}\right|$ is the molecular separation, and  $\hat{r}$ is a unit vector.

We introduce the geometric tensor $g_{\alpha\beta\left. \right| i,j}$  in a Cartesian coordinate system, relating $\boldsymbol{p}_{j}$ to $\boldsymbol{E}_{i,j}$ from  Eq. (\ref{eq:labframeE}),
\begin{equation}
g_{\alpha\beta \left.\right| i,j} = \displaystyle\left(3[\left(\hat{r}_{i}-\hat{r}_{j}\right)\cdot \hat{\alpha}] [\left(\hat{r}_{i}-\hat{r}_{j}\right) \cdot \hat{\beta} ] - \boldsymbol{\delta}_{\alpha\beta}\right) \displaystyle \frac{v_c} {\left|\boldsymbol{r}_{i} - \boldsymbol{r}_{j}\right|^{3}} ,
\label{eq:geofact}
\end{equation}
where the Greek subscripts represent the spatial Cartesian components and $\boldsymbol{\delta}_{\alpha\beta}$ is the Kronecker delta. Here, we have introduced a characteristic volume, $v_c$, which makes the geometric tensor dimensionless.

When the total field at molecule $i$ is sufficiently small, its dipole moment can be approximated as a power series,
\begin{eqnarray}
p_{\alpha\left.\right|i} &=& k_{\alpha\left.\right|i}^{\left(0\right)} + k_{\alpha\beta\left.\right|i}^{\left(1\right)} E_{\beta\left.\right|i} + k_{\alpha\beta\mu\left.\right|i}^{\left(2\right)} E_{\beta\left.\right|i} E_{\mu\left.\right|i} \nonumber \\
&+& k_{\alpha\beta\mu\nu\left.\right|i}^{\left(3\right)} E_{\beta\left.\right|i} E_{\mu\left.\right|i} E_{\nu\left.\right|i} + \cdots ,
\label{eq:polseries}
\end{eqnarray}
where $k_{i}^{\left(n\right)}$ is the $n$th-order polarizability of molecule $i$.

In a system of $N$ molecules, the total electric field is the vector sum of the applied field and the dipole fields due to all other molecules (higher-order multipole moments are ignored and we use a dipole approximation). Thus,
\begin{equation}
E_{\alpha\left.\right|i} = E_{\alpha\left.\right|i}^{a} + \sum_{j\neq i}^{N-1} E_{\alpha\left.\right|i,j}^{d} ,
\label{eq:totalE}
\end{equation}
where $E_{\alpha\left.\right|i}^{a}$ is the $\alpha$-component of the applied field at the $i$th molecule and $E_{\alpha\left.\right|i,j}^{d}$ is the $\alpha$-component of the dipole field at the $i$th molecule from the $j$th molecule. It is common to write the linear and nonlinear responses in terms of the macroscopic field. Because of the microscopic focus of this paper,  we have defined the effective polarizability and susceptibility in terms of the applied field, $\boldsymbol{E}^a$, where the depolarization field \cite{moroz09.01} and self-field \cite{crens08.01} are included in the summation of all other dipole contributions to the electric field.\cite{dawson12.01,*dawson12.01corr} Using linearity the dipole field
Eq. (\ref{eq:geofact}) can be written as
\begin{equation}
E_{\alpha\left.\right|i,j}^{d} = g_{\alpha\beta\left.\right|i,j} \displaystyle \frac{p_{\beta\left.\right|j}}{v_c} .
\label{eq:dipwgeofact}
\end{equation}

Because of the computational approach we use, it is natural to scale all $\left|\boldsymbol{p}_{j}\right|$ to $\left|\boldsymbol{p}_{i}\right|$ defining,
\begin{equation}
g_{\alpha\beta\left.\right|i}^{\left(N-1\right)} = \sum_{j\neq i}^{N-1} g_{\alpha\beta\left.\right|i,j} {\cal P}_{\beta\left.\right|i,j} ,
\label{eq:geotensorscale}
\end{equation}
where
\begin{equation}
{\cal P}_{\alpha\left.\right|i,j} = \frac{p_{\alpha\left.\right|j}} {p_{\alpha\left.\right|i}} .
\label{eq:scaletensorfactor}
\end{equation}

The factor $g_{\alpha\beta\left.\right|i}^{\left(N-1\right)}$ depends on the particular map of the \textit{a priori} molecular polarizations. A brute force approach would be to solve the set of polarization equations for every interacting molecule in the system. A simpler approach would be to approximate the value of ${\cal P}_{\alpha\left.\right|i,j}$ with an iterative method. Choosing the latter approach, we solve for $p_{\alpha\left.\right|i}^{\left[1\right]}$ in the equation
\begin{eqnarray}
p_{\alpha\left.\right|i}^{\left[1\right]} &=& k_{\alpha\left.\right|i}^{\left(0\right)} + k_{\alpha\beta\left.\right|i}^{\left(1\right)}\left(E_{\beta\left.\right|i}^{a} + \sum_{j\neq i}^{N-1} g_{\beta\gamma\left.\right|i,j} {\cal P}_{\gamma\left.\right|i,j}^{\left[0\right]} \displaystyle \frac{p_{\gamma\left.\right|i}^{\left[1\right]}}{v_c} \right) \nonumber \\
&+& k_{\alpha\beta\mu\left.\right|i}^{\left(2\right)}\left(E_{\beta\left.\right|i}^{a} + \sum_{j\neq i}^{N-1} g_{\beta\gamma\left.\right|i,j} {\cal P}_{\gamma\left.\right|i,j}^{\left[0\right]} \displaystyle \frac{p_{\gamma\left.\right|i}^{\left[1\right]}}{v_c} \right) \nonumber \\
&\times& \left(E_{\mu\left.\right|i}^{a} + \sum_{j\neq i}^{N-1} g_{\mu\nu\left.\right|i,j} {\cal P}_{\nu\left.\right|i,j}^{\left[0\right]} \displaystyle \frac{p_{\nu\left.\right|i}^{\left[1\right]}}{v_c} \right) + \cdots ,
\label{eq:iterapproach1}
\end{eqnarray}
where
\begin{equation}
{\cal P}_{\alpha\left. \right| i,j}^{\left[0\right]} = \frac{p_{\alpha\left.\right|j}^{\left[0\right]}} {p_{\alpha\left.\right|i}^{\left[0\right]}}
\label{eq:Pfracdefinition}
\end{equation}
and
\begin{equation}
p_{\alpha\left.\right|i}^{\left[0\right]} = k_{\alpha\left.\right|i}^{\left(0\right)} + k_{\alpha\beta\left.\right|i}^{\left(1\right)} E_{\beta\left.\right|i}^{a} + k_{\alpha\beta\mu\left.\right|i}^{\left(2\right)} E_{\beta\left.\right|i}^{a} E_{\mu\left.\right|i}^{a} + \cdots .
\label{eq:p0define}
\end{equation}

Then through an iterative method we solve for $p_{\alpha\left.\right|i}^{\left[n\right]}$ via the previously evaluated ${\cal P}_{\alpha\left.\right|i,j}^{\left[n-1\right]}$. The Appendix discusses the iterative process for higher-order corrections when a single iteration is not a sufficient approximation of ${\cal P}_{\alpha \left. \right| i,j}$.

Far from the strongly coupled regime, we approximate the interactions using only the first-order iterative correction to $g_{\alpha\beta\left.\right|i}^{\left(N-1\right)}$. We then define
\begin{equation}
f_{\alpha\beta\left.\right|i}^{\left(N-1\right)} = \sum_{j\neq i}^{N-1} g_{\alpha\beta\left.\right|i,j} {\cal P}_{\beta\left. \right| i,j}^{\left[0\right]} .
\label{eq:geotensorscale}
\end{equation}
Note that in this weakly coupled regime $f_{\alpha\beta\left.\right|i}^{\left(N-1\right)} \approx g_{\alpha\beta\left.\right|i}^{\left(N-1\right)}$ because the intermolecular responses are much less than every molecule's response to the applied field, \textit{i}.\textit{e}., when $k^{\left(1\right)}/r^3 \ll 1$. In addition, Eq. (\ref{eq:geotensorscale}) presupposes $E_{i}^{a} \neq 0$.

For dipole field distributions, ${\cal P}_{\alpha\left. \right| i,j}^{\left[0\right]}$ can be approximated by $E_{j}^{a}/E_{i}^{a}$ when $k^{\left(1\right)}E^{a} \gg k^{\left(n\right)}\left(E^{a}\right)^n$ for $n>1$; otherwise, the values of $k_{\alpha\beta\mu\nu\cdots}^{\left(n\right)}$ must be known to find a value for $f_{\alpha\beta\left.\right|i}^{\left(N-1\right)}$. We later use this approximation to generate dipole field maps. Again, for strongly interacting systems, higher-order corrections to the self-consistent equation, described in the Appendix, may be necessary for a more accurate approximation of $g_{\alpha\beta\left.\right|i}^{\left(N-1\right)}$.

For negligible second-order iterative corrections, substituting Eq. (\ref{eq:geotensorscale}) into Eq. (\ref{eq:iterapproach1}) gives
\begin{eqnarray}
p_{\alpha\left.\right|i} &\approx& k_{\alpha\left.\right|i}^{\left(0\right)} + k_{\alpha\beta\left.\right|i}^{\left(1\right)}\left(E_{\beta\left.\right|i}^{a} + f_{\beta\mu\left.\right|i}^{\left(N-1\right)} \displaystyle \frac{p_{\mu\left.\right|i}}{v_c}\right) \nonumber \\
&+& k_{\alpha\beta\mu\left.\right|i}^{\left(2\right)}\left(E_{\beta\left.\right|i}^{a} + f_{\beta\gamma\left.\right|i}^{\left(N-1\right)} \displaystyle \frac{p_{\gamma\left.\right|i}}{v_c}\right) \left(E_{\mu\left.\right|i}^{a} + f_{\mu\nu\left.\right|i}^{\left(N-1\right)} \displaystyle \frac{p_{\nu\left.\right|i}}{v_c}\right) \nonumber \\
&+& \cdots .
\label{eq:polseriesEd}
\end{eqnarray}
Using Eq.~(\ref{eq:polseriesEd}), we  solve for the effective (hyper)polarizabilities, where
\begin{equation}
k_{\mathrm{eff},\alpha\beta\mu\nu\cdots\left.\right|i}^{\left(n\right)} = \left. \frac{1}{n!}\frac{\partial^n p_{\alpha\left.\right|i}}{\partial E_{\alpha\left.\right|i}^{a} \partial E_{\beta\left.\right|i}^{a} \partial E_{\mu\left.\right|i}^{a} \partial E_{\nu\left.\right|i}^{a} \cdots}\right|_{\boldsymbol{E}_{i}^{a} = 0} .
\label{eq:knsolution}
\end{equation}

\subsection{Application to one-dimensional polarizable molecules}
\label{sub:onedim}

We eliminate the possibility of higher-order terms appearing in the lower-order effective (hyper)polarizabilities by assuming molecules with negligible permanent dipoles. Even for the spatially asymmetric systems we consider below, the associated static dipole ordering is typically orders of magnitude below that of the off resonant field induced effects we focus on. Note that this approximation still permits molecules having any higher-order response.\cite{ray04.01} Further, we assume that the only relevant tensor component is in the direction of the applied field. We also assume a lattice model.\cite{lebwo72.01,priez01.01,romano86.01} Although not strictly necessary, a lattice model allows for faster computation when simulating the dipolar field contributions.

Taking the applied field to be unidirectional and parallel to the $z$-axis, we reduce the tensor $f_{\alpha\beta\left.\right|i}^{\left(N-1\right)}$ to a vector $f_{\alpha z\left.\right|i}^{\left(N-1\right)}$. Because we are assuming a lattice model, we take the characteristic volume $v_c$ to be the volume of a unit cell, $v = \left|\boldsymbol{a} \cdot \left(\boldsymbol{b} \times \boldsymbol{c}\right)\right|$, where $\boldsymbol{a}$, $\boldsymbol{b}$, and $\boldsymbol{c}$ are the lattice vectors. Thus, the sum of the field contributions of all other molecules becomes
\begin{equation}
\sum_{j\neq i}^{N-1} E_{\alpha \left.\right| i,j}^{d} = f_{\alpha z \left. \right|i}^{\left(N-1\right)}\frac{p_{z\left.\right|i}}{v} .
\end{equation}
Note that the dimensionless geometric vector is scaled to the $i$th dipole that is induced by the applied field. Under these approximations we can now write a simplified equation for the induced dipole moment in the $z$-direction,
\begin{eqnarray}
p_{z \left. \right| i} &=& \sum_{n=1} k_{zz\cdots}^{\left(n\right)} \left(E_{i}^{a} + f_{zz \left. \right| i}^{\left(N-1\right)} \frac{p_{z \left. \right| i}}{v}  \right)^{n} . \label{eq:onedimz}
\end{eqnarray}

Again, in this section only, we have assumed that all dipoles polarize only along the applied field, and thus all tensor components other than $k_{zzz\cdots}^{\left( n \right)}$ are negligible. Although this model oversimplifies some scenarios that require the consideration of  molecular orientation (see Section \ref{sec:guesthost}), it allows for a single self-consistent equation, and evaluates the effective scalar (hyper)polarizabilities with respect to the applied field at each molecular site. Solving Eq. (\ref{eq:onedimz}) self-consistently for the dipole moment and substituting it into
\begin{equation}
k_{\mathrm{eff},i}^{\left(n\right)} = \left. \frac{1}{n!}\frac{\partial^n p_i} {\partial \left(E_{i}^{a}\right)^{n}}\right|_{E_{i}^{a} = 0} ,
\label{eq:knsolution}
\end{equation}
gives the effective scalar (hyper)polarizabilities in terms of the applied field. For example, a system of linearly polarizable molecules with no permanent dipole moment has an effective linear polarizability written as
\begin{equation}
k_{\mathrm{eff},i}^{\left(1\right)} = L_{i} k^{\left(1\right)} ,
\label{eq:paraalphatwodips}
\end{equation}
where the local field factor, $L_{i}$, at the $i$th molecule's location is given by
\begin{equation}
L_{i} = \left(1 - f_{zz \left. \right| i}^{\left(N-1\right)} \frac{k^{\left(1\right)}}{v}\right)^{-1} .
\label{eq:localfieldmodel}
\end{equation}
For a convergent solution everywhere, $k^{\left(1\right)} f_{zz \left. \right| i}^{\left(N-1\right)} < v$ for all $i$ molecules, otherwise the local field factor diverges.\cite{dawson11.02,dawson11.03} The average linear susceptibility is then written as
\begin{equation}
\left\langle\chi^{\left(1\right)}\right\rangle = \frac{1}{Nv} \displaystyle \sum_{i=1}^{N} k_{\mathrm{eff},i}^{\left(1\right)} .
\label{eq:linearsusCMrel}
\end{equation}
Here, $\chi$ is defined in terms of the applied field. Thus, Eq. (\ref{eq:localfieldmodel}) is analogous, but not equal, to the Lorentz-Lorenz local field factor.

\subsection{First-order corrections to microscopic cascading}
\label{sec:hyperpol}

Cascading lower-order nonlinearities to give higher-order nonlinear responses has been well understood and is inherent to the power series approximation of nonlinear optics.\cite{andre91.01,assan92.01,dolga12.01} The effective (hyper)polarizabilities are a combination of the highest-order response and cascaded lower-order responses. When near resonance, one must be careful to account for the imaginary (non-degenerate frequency mixing, absorption, etc.) and real (linear and nonlinear indices) components of the (hyper)polarizabilities. All tensor components are approximately real in the far off-resonant (below resonance) case to which we limit ourselves.

Taking into account only the largest contributing tensor component of the real molecular responses (the components purely in the direction of the applied field), and under the approximations in Section \ref{sub:onedim}, the first through sixth effective hyperpolarizabilities are
{\allowdisplaybreaks
\begin{eqnarray}
k_{\mathrm{eff},i}^{\left(2\right)} &=& L_{i}^{3} k^{\left(2\right)} , \label{eq:parabetatwodips} \\
k_{\mathrm{eff},i}^{\left(3\right)} &=& L_{i}^{4} k^{\left(3\right)} + 2 L_{i}^5 F_{i} \left(k^{\left(2\right)}\right)^2 , \label{eq:paragammatwodips} \\
k_{\mathrm{eff},i}^{\left(4\right)} &=& L_{i}^{5} k^{\left(4\right)} + 5 L_{i}^{6} F_{i} k^{\left(2\right)} k^{\left(3\right)} + 5 L_{i}^{7} F_{i}^{2} \left(k^{\left(2\right)}\right)^3  , \label{eq:paradeltatwodips} \\
k_{\mathrm{eff},i}^{\left(5\right)} &=& L_{i}^{6} k^{\left(5\right)} + 3 L_{i}^{7} F_{i} \left[\left(k^{\left(3\right)}\right)^2 + 2k^{\left(2\right)} k^{\left(4\right)}\right] \nonumber \\
&+& 21 L_{i}^{8} F_{i}^{2} \left(k^{\left(2\right)}\right)^2 k^{\left(3\right)} + 14 L_{i}^{9} F_{i}^{3} \left(k^{\left(2\right)}\right)^4 ,
\label{eq:paraepsilontwodips} \\
k_{\mathrm{eff},i}^{\left(6\right)} &=& L_{i}^{7} k^{\left(6\right)} + 7 L_{i}^{8} F_{i} \left[k^{\left(3\right)} k^{\left(4\right)} + k^{\left(2\right)} k^{\left(5\right)}\right] \nonumber \\
&+& 28 L_{i}^{9} F_{i}^{2} k^{\left(2\right)} \left[\left(k^{\left(3\right)}\right)^2 + k^{\left(2\right)} k^{\left(4\right)}\right] \nonumber \\
&+& 84 L_{i}^{10} F_{i}^{3} \left(k^{\left(2\right)}\right)^3 k^{\left(3\right)} + 42 L_{i}^{11} F_{i}^{4} \left(k^{\left(2\right)}\right)^5 , \label{eq:parazetatwodips} \\
k_{\mathrm{eff},i}^{\left(7\right)} &=& L_{i}^{8} k^{\left(7\right)} +4 L_{i}^{9} F_{i} \left[2 k^{\left(2\right)} k^{\left(6\right)} + 2 k^{\left(3\right)} k^{\left(5\right)} + \left(k^{\left(4\right)}\right)^2 \right] \nonumber \\
+ &12& L_{i}^{10} F_{i}^{2} \left[\left(k^{\left(3\right)}\right)^3 + 3 \left(k^{\left(2\right)}\right)^2 k^{\left(5\right)} + 6 k^{\left(2\right)} k^{\left(3\right)} k^{\left(4\right)}\right] \nonumber \\
&+& 60 L_{i}^{11} F_{i}^{3} \left(k^{\left(2\right)}\right)^2 \left[2 k^{\left(2\right)} k^{\left(4\right)} + 3 \left(k^{\left(3\right)}\right)^2 \right] \nonumber \\
&+& 330 L_{i}^{12} F_{i}^{4} \left(k^{\left(2\right)}\right)^4 k^{\left(3\right)} + 132 L_{i}^{13} F_{i}^{5} \left(k^{\left(2\right)}\right)^6 \label{eq:paraetatwodips} ,
\end{eqnarray}}
where
\begin{equation}
F_{i} = f_{zz \left. \right| i}^{\left(N-1\right)}\frac{k^{\left(1\right)}}{v} .
\label{eq:fieldfactorscomb}
\end{equation}
All lower-order terms ($k^{\left(0\right)}$ is assumed to be zero) in the nonlinear polarization series contribute to the higher-order hyperpolarizabilities in Eqs. (\ref{eq:parabetatwodips})-(\ref{eq:paraetatwodips}). The cascading contributions are ordered in terms of powers of $F_{i}$. For example, the mixing of two lower-order responses results in a higher-order response in which the magnitude depends on $F_{i}$, while the mixing of three lower-order responses depends on the value of $F_{i}^2$. Note that the dipole approximation may not be sufficient to express the effective response in many molecular systems because additional terms in the multipole expansion may make significant contributions to the effective hyperpolarizabilities.

\section{Applications to single-component systems}
\label{sec:appsing}

\subsection{Bound and strained systems}
\label{sub:boundstrained}

\begin{figure}[t]
\centering\includegraphics[scale=1]{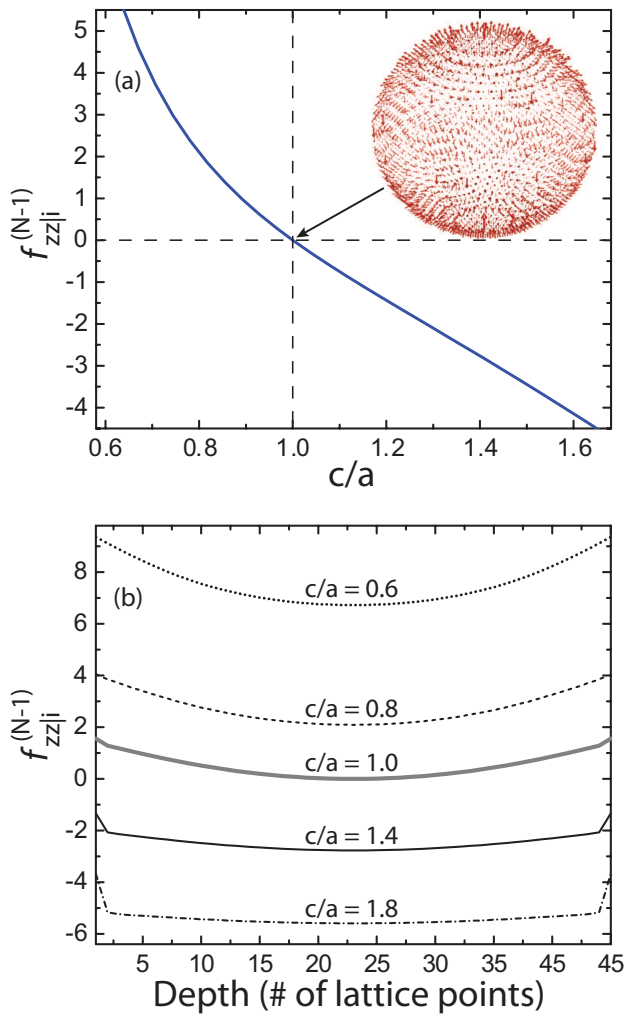}
\caption{(a) The first-order correction to $f_{zz \left. \right| i}^{\left(N-1\right)}$ for a molecule at the center of a sphere as a function of the lattice vector in the $z$-direction, $c$, divided by the lattice vector perpendicular to the field, $a$. The inset shows a sphere constructed from a cubic lattice where molecules near the surface have a nonzero $f_{\alpha z \left. \right| i}^{\left(N-1\right)}$ due to surface roughness. (b) The $zz$-component of the first-order approximation to the geometric factor, $f_{zz \left. \right| i}^{\left(N-1\right)}$, as a function of depth through the center of a strained $45\times45\times45$ cubic lattice (boundaries have zero electric flux).}
\label{fig:GeoFactCube}
\end{figure}

Among the geometric quantities affecting the susceptibility in a lattice with a finite number of atoms/molecules are the shape of the surface that contains the lattice, the shape of a primitive cell, and the incident beam (applied field) profile. Previous investigations for a top hat beam through a thin film \cite{dawson12.01,*dawson12.01corr} show enhancements due to cascading when a system is sharply bounded along the beam direction. The shape of the primitive cell is also known to change the local field in strained crystal lattices.\cite{herzf28.01,muell35.01,dunmu71.01,dunmu72.01,palff77.01}

There are many models that assume a potential from permanent dipoles on an infinite Bravais lattice for approximating macroscopic systems,\cite{born54.01,ewald12.01,ewald17.01} but we wish to approach the boundary problem via field-matter interactions, beginning with the perfect dipole approximation at each point on the lattice. This method requires knowledge about the entire system and all boundary locations, and, thus is more computationally expensive when calculating large systems.

We can also strain the lattice to change the cascaded nonlinear response of the system. Taking a large system of molecules on a tetragonal lattice with constants $\{a$,$a$,$c\}$, the $zz$-component of the geometric factor, $f_{zz \left. \right| i}^{\left(N-1\right)}$, monotonically increases as $c/a$ decreases. Figure \ref{fig:GeoFactCube}(a) shows how the z-component scales as a function of $c/a$ for a molecule located in the center of a large sphere constructed from tetragonal primitive cells.

\begin{figure*}[t]
\centering\includegraphics[scale=1]{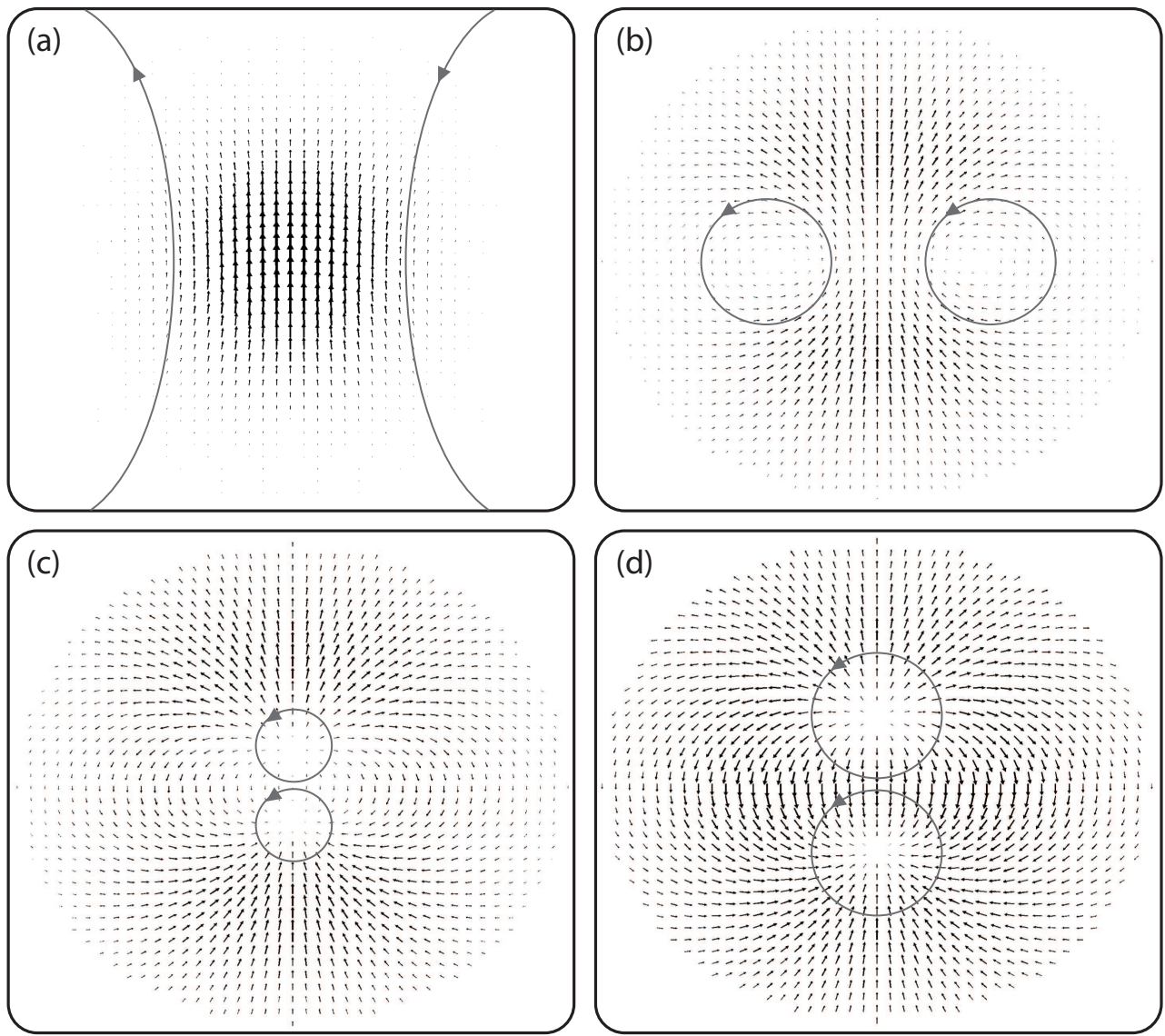}
\caption{A vector diagram of the first-order correction to the linear dipole fields for the center layer of a thin film subject to a vertically polarized Gaussian beam. The film thicknesses are (a) 1 layer, (b) 7 layers, (c) 15 layers, and (d) 55 layers of a cubic lattice system with side lengths that extend far beyond the edge of the graphic. Typical relevant line integrals noted in the text are explicitly shown.}
\label{fig:GaussCenter}
\end{figure*}

As anticipated, the dimensionless geometric factor, $f_{zz \left. \right| i}^{\left(N-1\right)}$, rapidly decreases and becomes negative as $c/a$ is increased due to the influence of \textit{all other} molecules. In contrast, $f_{zz \left. \right| i}^{\left(N-1\right)}$ rapidly becomes large as $c/a$ falls below unity. A defining feature appears when $c=a$, where all cascading fields for this center molecule cancel, \textit{i}.\textit{e}., $F_{i} = 0$.  Thus, for large cascading enhancements (large $f_{zz \left. \right| i}^{\left(N-1\right)}$), one would prefer aligned disk-like molecules with the applied field oriented along the short molecular axis as opposed to rod-like molecules with the applied field oriented along the long molecular axis.

By considering both microscopic structure and macroscopic geometry, we can further increase $F_{i}$ for systems of molecules with constant $v$. Figure \ref{fig:GeoFactCube}(b) shows how $f_{zz \left. \right| i}^{\left(N-1\right)}$ varies between two transverse interfacial boundaries in a strained $45\times45\times45$ cube with a tetragonal lattice structure. At the boundaries, even in a highly-elongated tetragonal lattice, $f_{zz \left. \right| i}^{\left(N-1\right)}$ is much larger than the next calculated interior-location.

\subsection{Dipolar electric field distributions}
\label{sub:dipolefield}

This section examines fixed lattices subject to an optical beam profile that is smaller than the transverse size of the system. An example would be that of the previously studied top hat beam,\cite{dawson12.01,*dawson12.01corr} where the molecules both inside and outside the beam are optically relevant. Here, we focus our attention on a long-wavelength monochromatic beam with a Gaussian profile.

Figure \ref{fig:GaussCenter} shows four vector diagrams of the directional components of the field due to the polarization of all other molecules. The unit cells are cubic and the size of the arrows are relative to each other in all parts (a)-(d). In these diagrams, we plot only the first-order iterative correction to a vertically polarized applied field. The diagrams show the induced field at the center layer of a thin film, where we have truncated the illustrations beyond the edge of the beam waist (where the electric field falls below $1/e$ of the peak value). Note that we assume that the beam is unchanged during transport through the film's thickness, but we expect  that the longitudinal propagation through thick films will be affected by the assumed nonlinear index via the self-focusing phenomenon and the inhomogeneous cascading predicted by $f_{\alpha z \left. \right| i}^{\left(N-1\right)}$.

As shown in the progression from Fig. \ref{fig:GaussCenter}(a)-(d), the competition between the \textit{in-plane} and \textit{out-of-plane} dipoles contribute to the electric field in the middle layer in nontrivial ways. For samples thicker than 55-layers, there is little change in the field profile at the center layer for these lattice/beam parameters. In the scaling limit the topology of the first order correction to the field profile at the center layer depends only on the ratio of the beam diameter to the thickness.

\begin{center}
\begin{table*}[t!]
\caption{Monolayers of C$_{60}$ subject to a vertically polarized Gaussian beam. $\left\langle \chi_{\mathrm{casc}}^{\left(5\right)} \right\rangle$ values are $\times 10^{-26}$cm$^{4}$erg$^{-2}$.}
\begin{minipage}{1\textwidth}
\centering
\begin{tabular}{| c | c | c | c | c |}
\hline
Lattice type & \hspace{.05cm} Vertical hexagonal \hspace{.05cm} & \hspace{.05cm} Horizontal hexagonal \hspace{.05cm} & \hspace{.05cm} Vertical honeycomb \hspace{.05cm} & \hspace{.05cm} Horizontal honeycomb \hspace{.05cm} \\[2.5pt]
\hline
\raisebox{9.5ex}{Diagram} & \raisebox{-1.5ex}{\includegraphics{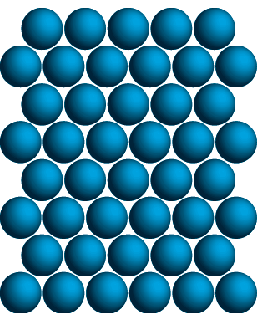}} & \raisebox{-1.5ex}{\includegraphics{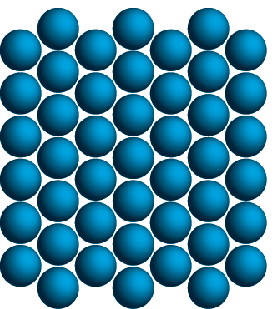}} & \raisebox{-1.5ex}{\includegraphics{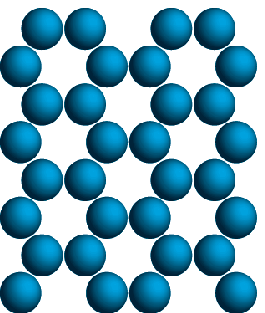}} & \raisebox{-1.5ex}{\includegraphics{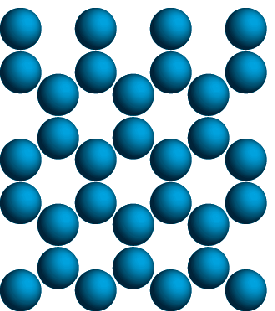}} \\
\hline
\hspace{.15cm} $\left\langle \chi_{\mathrm{casc}}^{\left(5\right)} \right\rangle^{\left[3\right]}$ \hspace{.1cm} & 4.34 & 4.34 & 1.91 & 1.91 \\[2.5pt]
\hline
\hspace{.15cm} $\left\langle \left|p_y\right|^{\left[1\right]}\right\rangle$ / $\left\langle \left|p_z\right|^{\left[1\right]} \right\rangle$ \hspace{.1cm} & 0.0058 & 0.0006 & 0.0011 & 0.0006 \\[2.5pt]
\hline
\end{tabular}
\end{minipage}
\begin{minipage}{1\textwidth}
\begin{tabular}{ l }
The difference between the response in a hexagonal and honeycomb lattice can be understood by using a $2/3$ fill \\[-6.5pt] factor in Eq. (\ref{eq:chi5casc}). The lattice geometry and nearest neighbor distance remain the same, but the concentration has \\[-6.5pt] been reduced by the fill factor. Therefore, we find that the computed honeycomb response is roughly $4/9$ that of \\[-6.5pt] the computed hexagonal lattice, confirming the greater significance of cascading in the filled, close-packed, structure.
\end{tabular}
\end{minipage}
\label{table:hexC60}
\end{table*}
\end{center}

Topological considerations are useful for understanding the successive frames as one adds layers, where we adjoin a ``neighborhood at infinity'' to make each of these a map of a vector field on a spherical surface, $S^2$. Imagining the vector field over the sphere, there are two zeros of the vector fields in each panel of Fig. \ref{fig:GaussCenter}. The line integrals of the vector fields around regions containing the zeros in Figs.  \ref{fig:GaussCenter}(a) and \ref{fig:GaussCenter}(b)  have matched positive and negative vorticity. On the other hand, the line integrals of the Hodge dual vector field on those two diagrams are zero.\cite{goldb62.01} The opposite is true for Figs. \ref{fig:GaussCenter}(c) and \ref{fig:GaussCenter}(d), where the line integrals of the Hodge dual give positive and negative vorticity around the zeros of the field. Indeed, Figs. \ref{fig:GaussCenter}(b) and \ref{fig:GaussCenter}(d) possess the same topological features as each other's Hodge duals just as vortices and sources are Hodge duals. In this vein, there exists a diagram between Figs. \ref{fig:GaussCenter}(b) and \ref{fig:GaussCenter}(c) that is nearly self-dual.

\subsection{Real systems: monolayers of close-packed C$_{60}$}
\label{sub:C60}

We now focus our attention on monolayers of close-packed C$_{60}$ in different lattice structures illuminated by a coherent beam. We chose C$_{60}$ due to large third-order susceptibility. Because there are larger cascading enhancements at higher concentration, we consider a hexagonal close-packed structure with a center-of-mass separation distance of approximately $10.04{\AA}$.\cite{gimze94.01} Note that a past study showed that perturbed energy states from cascaded molecules have small effects on large off-resonant nonlinear optical responses,\cite{dawson11.03} where we only observe significant effects after the molecules have passed into the ``forbidden'' zone in which they spatially overlap.

Due to the large intrinsic values of the odd-ordered susceptibilities of C$_{60}$, the polarizability and second hyperpolarizability are estimated by the three level ansatz.\cite{kuzyk05.02,kuzyk06.01,perez08.01} The values for the oscillator strengths and their corresponding transition energies were previously reported by Leach, \textit{et al}.\cite{leach92.01,bulga92.01} Truncating the (hyper)polarizabilities to only three states gives $k^{\left(1\right)} = 1.85\times10^{-23}$cm$^{3}$ and $k^{\left(3\right)} = 3.41\times10^{-35}$erg$^{-1}$cm$^{5}$.\cite{dawson12.01,*dawson12.01corr} Also, $k^{\left(0\right)} \approx 0$, $k^{\left(2\right)} \approx 0$, and $k^{\left(4\right)} \approx 0$ due to the near spherical symmetry of C$_{60}$. Note that using the standard time-dependent perturbation approach,\cite{orr71.01} truncation to a three level model may greatly overestimate the higher-order polarizabilities.

As a comparative study, we look at the vertical and horizontal orientations of the lattice as well as a honeycomb structure. The off-resonant beam carrying the applied field propagates in the $x$-direction and vertically polarized in the $z$-direction. The diameter of the Gaussian beam is $150\,$nm, where the location of the electric field is $1/e$ of its peak value. The calculated region for all contributions from molecular interactions has a diameter of $180\,$nm. The average effective susceptibilities are calculated within the beam waist after all contributions from the extended region have been taken into account.

We consider only the scalar (hyper)polarizabilities, though small values of $p_{y\left. \right| i}$ will be present, and we calculate out to a third-order iteration. The average effective fifth-order susceptibility (susceptibility defined by the applied field with cascading enhancements) for the region inside the beam waist is
\begin{equation}
\left\langle \chi^{\left(5\right)}\right\rangle^{\left[n\right]} = \frac{k^{\left(5\right)}}{V} \sum_{i}^{N} \left(L_{i}^{\left[n\right]}\right)^{6} + \left\langle \chi_{\mathrm{casc}}^{\left(5\right)} \right\rangle^{\left[n\right]}
\label{eq:chi5}
\end{equation}
where
\begin{equation}
\left\langle \chi_{\mathrm{casc}}^{\left(5\right)} \right\rangle^{\left[n\right]} = 3 N \left(\frac{k^{\left(3\right)}}{V}\right)^2 \sum_{i}^{N} \left(L_{i}^{\left[n\right]}\right)^{7} \left(f_{i}^{\left(N-1\right)}\right)^{\left[n\right]} .
\label{eq:chi5casc}
\end{equation}
Here, we denote the total volume by $V = Nv$ and the $n$th-order iterative correction by the superscript $^{\left[n\right]}$. The value of $\left\langle \chi_{\mathrm{casc}}^{\left(5\right)} \right\rangle$ is calculated from an arithmetic average. Table \ref{table:hexC60} lists values of $\left\langle \chi_{\mathrm{casc}}^{\left(5\right)} \right\rangle$ for the vertical and horizontal lattice alignments of hexagonal and honeycomb monolayers. The values of $\left\langle \chi_{\mathrm{casc}}^{\left(5\right)} \right\rangle^{\left[n\right]}$ for a Gaussian and top hat beam are similar even though the Gaussian beam has a smaller applied field at all molecules except at the center. This can be understood by the on-average increase of ${\cal P}_{i,j}$ as we move further from the center of the Gaussian beam. Note that although the responses between the two types of beam profiles are the same, the magnitude of the cascading contribution for a Gaussian beam (peak value of $10^6\,$StatV/cm for Gaussian and top hat beams) is smaller than that resulting from a top hat beam because the susceptibility is multiplied by the tapered Gaussian beam's applied field. Rotating the polarization of a linearly polarized beam between the vertical and horizontal lattice alignments also shows negligible changes in the cascading contribution.

For the hexagonal monolayer subject to a Gaussian beam profile, the values from the first and second iteration change by $<3\%$. Thus, a first-order approximation to the iterative method is fairly accurate in this scenario and does not carry the computational expense of higher orders that require interactions between polarization directions via tensor components. The iterative method converges quickly, typically changing only in the fifth digit from the second to the third iteration for these monolayers. All iterations after the second (tested out to 20 iterations for stability) showed a stable precision much greater than the uncertainties of the model due to the many approximations (point dipoles, truncated eigenstates, lattice precision, \textit{etc}.).

\section{APPROXIMATING CASCADING IN POLED GUEST-HOST SYSTEMS}
\label{sec:guesthost}

So far we have only considered systems with a single species of atom/molecule. The lattice model, however, can be further generalized to include several molecules with different optical properties. A dipole moment can then be written for individual molecules, where the dependencies on all fields are taken into account including the field contributions from the other species.

To illustrate the inclusion of more than one type of atom or molecule, we choose a dye-doped polymer system. The two main advantages of placing active nonlinear molecules in a polymer are (1) the large linear susceptibilities of many polymers that increase the local field and (2) the ability to align the nonlinear dopant in the medium.\cite{singe88.01,eich89.01,hamps90.01} We use the lattice approximation to model the field enhancement via a randomized occupation of the lattice sites by the guest species. A host cluster is approximated as a point dipole at each occupied lattice site, which we call the host cluster dipole approximation for non-conjugated polymers. The guest species is assumed to be uniaxially aligned, although this simplification may be removed for a more general result.

We assume that all dopant molecules in the system have (hyper)polarizabilities in the $z$-direction that are equal to the orientational averaged (hyper)polarizabilities and all other components are negligible, \textit{e.g.}, $\left\langle k^{\left(2\right)} \right\rangle = \left\langle k_{zzz}^{\left(2\right)} \right\rangle = \left\langle \cos^3 \theta \right\rangle k_{zzz}^{\left(2\right)}$ and $\left\langle k_{ijk}^{\left(2\right)} \right\rangle \approx 0$ for all cases other than $i=j=k=z$. This approximation is valid for one-dimensional molecules oriented at small angles from the direction of the electric field, where $\left\langle k_{zxx}^{\left(2\right)} \right\rangle = \left\langle k_{xzx}^{\left(2\right)} \right\rangle = \left\langle k_{xxz}^{\left(2\right)} \right\rangle = \left\langle \cos \theta \sin^2 \theta \right\rangle k_{zzz}^{\left(2\right)} / 2$, which is small due to the $\sin^2 \theta$ contribution.\cite{rasin85.01,kuzyk89.03} A full treatment of the cascading contributions to $k^{\left(3\right)}$ for a pair of one-dimensional molecules in an electric field at fixed locations is given in Ref. \cite{dawson11.03}. For our current example, however, we ignore the azimuthal angle and treat only the average polar angle in an attempt to reduce  orientational complexities. Therefore, for fixed molecules, we define $\kappa^{\left(n\right)} = \left\langle \cos^{n+1} \left(\theta\right) \right\rangle k^{\left(n\right)}$.

We define $p^A$ as the dipole moment of the linear host species and $p^B$ as the dipole moment of the guest species. The two dipole moment equations are
{\allowdisplaybreaks \begin{eqnarray}
p_{i}^A &=& \kappa_{A}^{\left(1\right)} \left(E_{i}^{a} + \displaystyle\sum_{j\neq i}^{N_A-1} h_{i,j} \frac{p_{j}^{A}}{v} + \displaystyle \sum_{j}^{N_B} f_{i,j} \frac{p_{j}^{B}}{v}\right) , \label{eq:firstdipA} \\
p_{i}^B &=& \displaystyle\sum_n \kappa_{B}^{\left(n\right)} \left(E_{i}^{a} + \displaystyle\sum_{j}^{N_A} h_{i,j} \frac{p_{j}^{A}}{v} + \displaystyle\sum_{j\neq i}^{N_B-1} f_{i,j} \frac{p_{j}^{B}}{v}\right)^n  \label{eq:seconddipB}
\end{eqnarray}}
where $h_{i,j}$ and $f_{i,j}$ are the geometry-dependent factors (to a first-order iterative approximation) for species $A$ and $B$ that account for a dipole field from all $j$ molecules. We treat $h_{i,j}$ and $f_{i,j}$ as scalars because we choose our uniaxial molecules to be aligned with the applied field's polarization. Note that for higher-order iterative corrections, we must keep track of each host molecule's $\left(\kappa_{A\left.\right|i}^{\left(1\right)}\right)^{\left[m\right]}$ and guest molecule's $\left(\kappa_{B\left.\right|i}^{\left(n\right)}\right)^{\left[m\right]}$.

Solving Eq. (\ref{eq:firstdipA}) gives
\begin{equation}
p_{i}^{A} = \kappa_{A}^{\left(1\right)} {\cal L}_{i}^{A} \left(E_{i}^{a} + \displaystyle\sum_{j}^{N_B} f_{i,j} \displaystyle\frac{p_{j}^{B}}{v}\right) ,
\label{eq:solvedfordipA}
\end{equation}
where
\begin{equation}
{\cal L}_{i}^{A}  = \left(1-\displaystyle\sum_{j\neq i}^{N_A-1} h_{i,j} {\cal P}_{i,j}^{A} \frac{\kappa_{A}^{\left(1\right)}}{v}\right)^{-1} .
\label{eq:calLi}
\end{equation}
Here, ${\cal L}_{i}^{A}$ is the first-order correction to the field at a host cluster due to all other host clusters, where we have also included the scaling factor ${\cal P}_{i,j}^{A} = p_{j}^{A}/p_{i}^{A}$ for molecules subject to a spatially varying applied field with the same approximations described in Section \ref{sec:firstorder}.

Substituting Eq. (\ref{eq:solvedfordipA}) into Eq. (\ref{eq:seconddipB}) gives
\begin{eqnarray}
p_{i}^B = \displaystyle\sum_n \kappa_{B}^{\left(n\right)} \left[\left(1 + {\cal Q}_i\right) E_{i}^{a} + \left({\cal S}_i + f_{i}^{\left(N_B -1 \right)} \right) \frac{p_{i}^{B}}{v}\right]^n
\label{eq:ABsolve}
\end{eqnarray}
where
\begin{eqnarray}
{\cal Q}_i &=& \frac{\kappa_{A}^{\left(1\right)}}{v}\displaystyle\sum_{j}^{N_A} h_{i,j} {\cal L}_{j}^{A} {\cal E}_{i,j} , \label{eq:Qidef} \\
{\cal S}_i &=& \frac{\kappa_{A}^{\left(1\right)}}{v} \displaystyle\sum_{j}^{N_A} h_{i,j} {\cal L}_{j}^{A} \displaystyle\sum_{k}^{N_B} f_{j,k} {\cal P}_{i,k}^{B} , \label{eq:Sidef}
\end{eqnarray}
and
\begin{equation}
f_{i}^{\left(N_B-1\right)} = \displaystyle\sum_{j\neq i}^{N_B-1} f_{i,j} {\cal P}_{i,j}^{B} .
\label{eq:fiNm1B}
\end{equation}
The last term in Eq. (\ref{eq:Qidef}) has a direct dependence on a spatially varying applied field, where
\begin{equation}
{\cal E}_{i,j} = \displaystyle\frac{E_{j}^{a}}{E_{i}^{a}} .
\label{eq:Efracdef}
\end{equation}
Solving Eq. (\ref{eq:ABsolve}) self-consistently and substituting into Eq. (\ref{eq:knsolution}) gives the (hyper)polarizabilities of guest molecules. Off-resonance, the first-order contributions to the first three effective polarizabilities of the $i$th guest molecule are
\begin{eqnarray}
\kappa_{\mathrm{eff},B,i}^{\left(1\right)} &=& \kappa_{B}^{\left(1\right)} \displaystyle \frac{1+{\cal Q}_i}{1-\left(f_{i}^{\left(N_B-1\right)}+{\cal S}_i \right) \displaystyle\frac{\kappa_{B}^{\left(1\right)}} {v}} , \label{eq:kapB1} \\
\kappa_{\mathrm{eff},B,i}^{\left(2\right)} &=& \kappa_{B}^{\left(2\right)} \displaystyle \frac{\left(1+{\cal Q}_i\right)^2}{\left(1-\left(f_{i}^{\left(N_B-1\right)}+{\cal S}_i \right) \displaystyle\frac{\kappa_{B}^{\left(1\right)}} {v} \right)^3} , \label{eq:kapB2}
\end{eqnarray}
and
{\allowdisplaybreaks
\begin{eqnarray}
\kappa_{\mathrm{eff},B,i}^{\left(3\right)} &=& \displaystyle \frac{\kappa_{B}^{\left(3\right)} \left(1+{\cal Q}_i\right)^3}{\left(1-\left(f_{i}^{\left(N_B-1\right)}+{\cal S}_i \right) \displaystyle\frac{\kappa_{B}^{\left(1\right)}} {v} \right)^4} \label{eq:kapB3} \\
&+& \displaystyle \frac{2}{v} \left(\kappa_{B}^{\left(2\right)}\right)^2 \displaystyle \frac{\left(1+{\cal Q}_i\right)^3 \left(f_{i}^{\left(N_B-1\right)}+{\cal S}_i \right)} {\left(1-\left(f_{i}^{\left(N_B-1\right)}+{\cal S}_i \right) \displaystyle \frac{\kappa_{B}^{\left(1\right)}} {v} \right)^5} . \nonumber
\end{eqnarray}}

Equations (\ref{eq:kapB1})-(\ref{eq:kapB3}) are similar in form to Eqs. (\ref{eq:paraalphatwodips}), (\ref{eq:parabetatwodips}), and (\ref{eq:paragammatwodips}) except for the terms ${\cal Q}_i$ and ${\cal S}_i$. The first additional term, ${\cal Q}_i$, comes from the self-consistent linear field correction to the guest molecules from the surrounding host material. The second additional term, ${\cal S}_i$, is similar to a second-order iterative correction in the single species model, where a nonlinear process from a guest molecule alters the field that a host cluster experiences (including field corrections from the host), which in turn affects the field at any guest molecule.

As an example, we consider a \textit{thin, poled, guest-host film of disperse orange 3 (DO3) molecules dissolved in poly(methyl methacrylate) (PMMA)}. DO3 is an azobenzene dye with a molecular weight of approximately $242\,$g/mol. PMMA has a density of $1.12\,$g/cm$^3$, and setting the cubic lattice constant to approximately $7.11\,$\AA (the volume of a cubic cell is $3.19\times10^{-22}\,$cm$^3$) gives an effective molecular weight of the host cluster to be that of DO3 (not the actual molecular weight of a host molecule). For PMMA with a dielectric constant, $\epsilon_r$, of $2.85$, we find a host cluster polarizability of approximately $3.27\times10^{-23}\,$cm$^3$ in Gaussian units via the Clausius-Mossotti equation for an isotropic material,
\begin{equation}
k_{A}^{\left(1\right)} = \displaystyle\frac{3 V}{4\pi N}\left(\frac{\epsilon_r -1}{\epsilon_r +2}\right) ,
\label{eq:claussmoss}
\end{equation}
where $V$ is the total volume given in units of cm$^3$, $N$ is the the number of host clusters, and the presence of $4\pi$ in the denominator (lack of $\epsilon_0$ in the numerator) converts the polarizability to Gaussian units. Note that $k_{A}^{\left(1\right)}$ is assumed to be isotropic, and therefore, $\kappa_{A}^{\left(1\right)} = k_{A}^{\left(1\right)}$. The guest molecules are assumed to be at a concentration of $1.56$\%, which roughly corresponds to one guest molecule per every $64$ lattice sites. We consider a sample of thickness $9.24\,$nm (13 lattice sites thick), subject to a Gaussian beam with a diameter of approximately $150\,$nm.

The guest molecules have an average polar angle of $\left\langle \theta \right\rangle = 15^\circ$, from the polarization orientation. The real off-resonant polarizability was evaluated using the ORCA program system \cite{neese12.01} and was $7.99\times 10^{-23}\,$cm$^{3}$. Here, we used the BP functional in conjunction with the TZV basis set.\cite{perde86.01,becke88.01,shaef92.01} The first and second hyperpolarizability of DO3 have been tabulated as $2.77\times 10^{-29}\,$erg$^{-1/2}$cm$^{4}$ and $2.56\times 10^{-34}\,$erg$^{-1}$cm$^{5}$, respectively.\cite{xiang01.01,audeb03.01}
The corresponding orientational averaged values at $15^\circ$ are $\kappa_{B}^{\left(1\right)} = 7.45\times 10^{-23}\,$cm$^{3}$, $\kappa_{B}^{\left(2\right)} = 2.50\times 10^{-29}\,$erg$^{-1/2}$cm$^{4}$, and $\kappa_{B}^{\left(3\right)} = 2.23\times 10^{-34}\,$erg$^{-1}$cm$^{5}$.

The average of the first-order iterative cascaded contribution to the orientationally averaged, scalar, second hyperpolarizability, $\left\langle\kappa_{\mathrm{eff}}^{\left(3\right)}\right\rangle$, was calculated to be $2.77\times 10^{-33}\,$erg$^{-1}$cm$^5$, where the first term in Eq. (\ref{eq:kapB3}) is $2.61\times 10^{-33}\,$erg$^{-1}$cm$^5$ and the second term in Eq. (\ref{eq:kapB3}) is $1.61\times 10^{-34}\,$erg$^{-1}$cm$^5$. The average third-order susceptibility as a function of the applied field (assuming a negligible nonlinear response of PMMA), $\left\langle \chi_{\mathrm{eff}}^{\left(3\right)} \right\rangle$, is $1.20\times 10^{-13}\,$erg$^{-1}$cm$^2$.  For comparative purposes, if we were to remove the PMMA and observe the DO3 in a gas phase while keeping the long molecular axis aligned with the field making an average polar angle of $15^\circ$, we calculate the orientationally averaged third-order susceptibility  $\left\langle\chi_{\mathrm{eff}}^{\left(3\right)}\right\rangle_{\mathrm{gas}-\mathrm{orient}} \approx N_B \kappa^{\left(3\right)}/ V$ to be $9.69\times 10^{-15}\,$erg$^{-1}$cm$^2$. Thus, the presence of the linear host greatly enhances the nonlinear susceptibility of the system due to both dipole field and cascading effects.

When modeling guest-host systems, the ${\cal Q}_i$'s depend on the details of the microscopic configuration. Note that these ${\cal Q}_i$'s, which refer to the host's linear modification to the applied field, can be both positive and negative. Our approach allows one to calculate the cascaded contribution \textit{ab} \textit{initio} for a nano-engineered system with a specific geometry.

\section{CONCLUSIONS}
\label{sec:conc}

We used a self-consistent method to derive the scalar, effective hyperpolarizabilities of bounded systems out to sixth-order. The lattice model allows for fast calculations of geometric factors that epitomize the electronic interactions between polarizable atoms/molecules. By substituting these geometric factors into the calculation for the response of a system, we have shown that boundary effects from thin films and deviations from a cubic lattice enhance the field at molecular locations and enhance the cascading contributions to the off-resonant optical responses. The resultant field due to dipoles induced by a Gaussian beam has been characterized for different film thicknesses. We have shown how in-plane and out-of-plane interaction affect the dipole field in these films for a Gaussian beam, and we have given a method to calculate these effects for other beam types. We also applied our approach to cascading to calculate the nonlinear cascaded contribution to the fifth-order susceptibility in monolayers of C$_{60}$. We found that with periodic monolayers, the cascading enhancement is directly related to the fill factor in the scaling limit.

We further developed this approach in application to a guest-host model, where a linear-optical host is doped with nonlinear-optical molecules. By limiting the study to fixed molecules, we derived expressions for the effective, nonlinear responses of the guest molecules that include all linear- and nonlinear-optical cascading configurations. We used a 1.56\% DO3-doped PMMA system as an example in which we show more than an order-of-magnitude increase in the third-order susceptibility with respect to an oriented gas state (no host present). This calculation showed how a thin film, even at small concentrations of nonlinear dopants, has a large impact on the nonlinear response. We derived an expression with a familiar local field factor, and also showed an additional factor that scales in powers with the response. This method does have current shortcomings such as lacking the inclusion of an on-resonant response and higher-order multipole moments. Future areas of improvement include beam profile deformation calculations while propagating through a material, calculations with higher resolution molecules that are not point-like, and methods for decreasing the computation time for larger systems.

\acknowledgements

The authors are grateful to the National Science Foundation for financial support from the Science and Technology Center for Layered Polymeric Systems under grant number DMR 0423914. The authors are also grateful for the support of the Department of Development, State of Ohio, the Chancellor of the Board of Regents, State of Ohio, and the Third Frontier Commission, which provided funding in support of the Research Cluster on Surfaces in Advanced Materials.

\appendix

\section*{Appendix A: Higher-order corrections and the iterative process}
\label{app:higherorder}
\setcounter{equation}{0}
\renewcommand{\theequation}{A{\arabic{equation}}}

For many systems, Eq. (\ref{eq:geotensorscale}) may not give a close enough approximation to the effective (hyper)polarizabilities. In these cases, further iterations to the self-consistent dipole equation are necessary to give a more accurate description of the off-resonant cascading contribution. For the first-order correction, we obtained solutions in terms of $f_{\alpha\beta \left. \right| i}^{\left(N-1\right)}$. The iterative method is described following Eq. (\ref{eq:iterapproach1}), and for the second-order correction, gives
{\allowdisplaybreaks \begin{eqnarray}
p_{\alpha\left.\right|i}^{\left[2\right]} &=& k_{\alpha\left.\right|i}^{\left(0\right)} + k_{\alpha\beta\left.\right|i}^{\left(1\right)}\left(E_{\beta\left.\right|i}^{a} + \sum_{j\neq i}^{N-1} g_{\beta\gamma\left.\right|i,j} {\cal P}_{\gamma\left.\right|i,j}^{\left[1\right]} \displaystyle \frac{p_{\gamma\left.\right|i}^{\left[2\right]}}{v_c} \right) \nonumber \\
&+& k_{\alpha\beta\mu\left.\right|i}^{\left(2\right)}\left(E_{\beta\left.\right|i}^{a} + \sum_{j\neq i}^{N-1} g_{\beta\gamma\left.\right|i,j} {\cal P}_{\gamma\left.\right|i,j}^{\left[1\right]} \displaystyle \frac{p_{\gamma\left.\right|i}^{\left[2\right]}}{v_c} \right) \nonumber \\
&\times& \left(E_{\mu\left.\right|i}^{a} + \sum_{j\neq i}^{N-1} g_{\mu\nu\left.\right|i,j} {\cal P}_{\nu\left.\right|i,j}^{\left[1\right]} \displaystyle \frac{p_{\nu\left.\right|i}^{\left[2\right]}}{v_c} \right) + \cdots ,
\label{eq:iterapproach2}
\end{eqnarray}}

The effective (hyper)polarizabilities given in Eqs. (\ref{eq:paraalphatwodips}) and (\ref{eq:parabetatwodips})-(\ref{eq:paraetatwodips}) are first-order corrections to the response of molecules that are polarized along the direction of the applied field. With a more rigorous approach, one can find the effective (hyper)polarizabilities for all possible components. Thus, we can define ${\cal P}_{\alpha\left. \right| i,j}^{\left[1\right]}$ in terms of these first-order effective (hyper)polarizabilities and the applied electric field,
\begin{equation}
{\cal P}_{\alpha\left. \right| i,j}^{\left[1\right]} = \displaystyle \frac{\displaystyle \sum_{n} \left(k_{\alpha \beta \mu \nu \cdots \left. \right|j}^{\left(n\right)}\right)^{\left[1\right]} E_{\beta \left. \right|j}^{a} E_{\mu \left. \right|j}^{a} E_{\nu \left. \right|j}^{a} \cdots } {\displaystyle \sum_{n} \left(k_{\alpha \beta \mu \nu \cdots \left. \right|i}^{\left(n\right)}\right)^{\left[1\right]} E_{\beta \left. \right|i}^{a} E_{\mu \left. \right|i}^{a} E_{\nu \left. \right|i}^{a} \cdots } ,
\label{eq:gencalP1A}
\end{equation}
where we have altered the notation of the effective (hyper)polarizabilities to account for higher-order corrections, \textit{i}.\textit{e}., $k_{\mathrm{eff}}$ in Section \ref{sec:firstorder} is the first-order correction in the iterative method $k^{\left[1\right]}$.

For a known spatial distribution of the applied electric field, Eq. (\ref{eq:gencalP1A}) has some specified value for each component of a molecule $i$ with respect to some other molecule $j$, which is similar to that used for the updates in Ref. \cite{kuhn08.01}. Once the single molecule (hyper)polarizabilities have been inserted into the first-order correction to the effective (hyper)polarizabilities, the first-order corrected effective (hyper)polarizabilities are inserted into Eq. (\ref{eq:gencalP1A}). Then, we define
\begin{equation}
\left(f_{\alpha\beta\left.\right|i}^{\left(N-1\right)}\right)^{\left[1\right]} = \sum_{j\neq i}^{N-1} g_{\alpha\beta\left.\right|i,j} {\cal P}_{\beta\left. \right| i,j}^{\left[1\right]} ,
\label{eq:geotensorscale1}
\end{equation}
\vspace{0.05cm}where $\left(f_{\alpha\beta\left.\right|i}^{\left(N-1\right)}\right)^{\left[0\right]}$ is given in Eq. (\ref{eq:geotensorscale}).

To find the second correction to the off-resonant (hyper)polarizabilities for the $i$th molecule, we substitute Eq. (\ref{eq:geotensorscale}) into Eq. (\ref{eq:iterapproach2}), and then substitute the resultant equation into
\begin{equation}
\left(k_{\alpha\beta\mu\nu\cdots\left.\right|i}^{\left(n\right)}\right)^{\left[2\right]} = \left. \frac{1}{n!}\frac{\partial^n p_{\alpha\left.\right|i}^{\left[2\right]}}{\partial E_{\beta\left.\right|i}^{a} \partial E_{\mu\left.\right|i}^{a} \partial E_{\nu\left.\right|i}^{a} \cdots}\right|_{\boldsymbol{E}_{i}^{a} = 0} .
\label{eq:knsol2}
\end{equation}
These values are the second-order corrections to the (hyper)polarizabilities. This simple step-by-step iterative process may be used to evaluate these higher-order corrections, where a loop may be implemented until the effective hyperpolarizabilities converge. Third order-corrections are found via the next iteration, where we replace the superscripts $^{\left[2\right]}$ by the superscripts $^{\left[3\right]}$ and use values obtained in from the second-order corrections by replacing the superscripts $^{\left[1\right]}$ by the superscripts $^{\left[2\right]}$. Following this same principle, higher-order iterative approximations can be obtained.

{\hfill}{\hfill}{\hfill}{\hfill}{\hfill}


%

\end{document}